\documentclass[intlimits,twoside,a4paper]{article}
\usepackage[cp1251]{inputenc}
\usepackage{subcaption}
\usepackage[eqsecnum]{cmpj3}
\usepackage{bm}

\newcommand{\creat}[2]{#1_{#2}^\dagger(\vec{r},\tau)}
\newcommand{\distroy}[2]{#1_{#2}(\vec{r},\tau)}

\newcommand{\intr}{\int \rd^3r}
\newcommand{\intau}{\int\limits_{0}^{\hbar\beta}\rd\tau}

\newcommand{\pardt}{\dfrac{\partial }{\partial \tau}}
\newcommand{\pard}[2]{\dfrac{\partial #1}{\partial #2}}

\newcommand{\sech}{\text{sech}}
\issue{2021}{24}{1}{13101}
\doinumber{10.5488/CMP.24.13101}
\title[Theoretical study of I-V characteristics]%
{Theoretical study of I-V characteristics in a coupled long Josephson junctions based on magnesium diboride superconductor }
\author[S.P. Chimouriya, B.R. Ghimir, J.H. Kim]{S.P. Chimouriya\refaddr{label1,label2},
        B.R. Ghimire\refaddr{label2}, J.H. Kim\refaddr{label3}}
\addresses{
\addr{label1} Department of Physics, Kathmandu University, Dhulikhel, Kavre, Nepal
\addr{label2} Central Department of Physics, Tribhuvan University, Kirtipur, Kathmandu, Nepal
\addr{label3}College of Science and Engineering, University of Houston Clear Lake, TX, USA
}
\date{Received February 08, 2020, in final form July 14, 2020}
\begin{document}

\maketitle

\begin{abstract}
In the present work, the current-voltage (I-V) characteristics in a coupled long Josephson junction based on magnesium diboride are studied by  establishing a system of equations of phase differences of various inter- and intra-band channels starting from the microscopic Hamiltonian of the junction system and simplifying it through the phenomenological procedures such as action, partition function, Hubbard-Stratonovich transformation (bosonization), Grassmann integral, saddle-point approximation, Goldstone mode, phase dependent effective Lagrangian and, finally, Euler-Lagrange equation of motion. The system of equations are solved using finite difference approximation for which the solution of unperturbed sine-Gordon equation is taken as the initial condition. Neumann boundary condition is maintained at both the ends so that the fluxon is capable of reflecting from the end of the system. The phase dependent current is calculated for different tunnel voltage and averaged out over space and time. The current-voltage characteristics are almost linear at low voltage and  non-linear at higher voltage which indicates that the more complicated physical phenomena at this situation may occur. At some region of the characteristics, there exist a negative resistance which means that the junction system can be used in specific electronic devices such as oscillators, switches, memories etc. The non-linearity is also sensitive to the layer as well as to the junction thicknesses. Non-linearity occurs for lower voltage and for higher junction and layer thicknesses.
\end{abstract}

\section{Introduction}
Superconductivity of magnesium diboride (MgB$_2$) was discovered in 2001 with transition temperature of about 39 K \cite{nagamutsu2001}. Since its discovery as a superconductor, it has attracted the attention of the many researchers in the related fields because of its higher transition temperature than that of other metallic compounds. The two-gap nature of MgB$_2$ offers different types of physical phenomena which urges the researchers to work in the context of both theoretical and experimental prospects. The electronic structure of MgB$_2$ is similar to graphite which consists of honeycomb boron layers separated by magnesium layers~\cite{mazin2003}. The energy gaps are about $\Delta_1=2$~meV which corresponds to two $\pi$-bands and about $\Delta_2=7$~meV which corresponds to $\sigma$-band. The state of Cooper pair corresponds to the gaps and are designated by the order parameters: $\psi_1=\Delta_1\re^{\ri\theta_1}$ for the first gap and $\psi_2=\Delta_2\re^{\ri\theta_2}$ for the second gap. The internal degree of freedom is the inter-band phase difference $\theta(\vec{r},t)=\theta_1-\theta_2$ \cite{gurevich2003}. The degree of freedom can be increased by forming the stack of MgB$_2$ interlocked with the insulator such as SiO$_2$, Al$_2$O$_3$ etc., which is referred to as stacked Josephson junction. As a result,  the Cooper pairs tunnel through the junction and the inter-band as well as intra-band phase textures are quite complicated.

In the present work, we  derived a system of perturbed sine-Gordon equations for the coupled long Josephson junction. Starting from the microscopic BCS model Hamiltonian of the system and undertaking a number of steps of phenomenological path integral formalism, the phase dependent effective action and, hence, the effective Lagrangian density are derived. The system of equations of phase dynamics is derived by using Euler-Lagrange equation of motion to minimize the effective Langrangian density. The system of equations are then solved numerically using the finite difference approximation imposing the Neumann boundary condition. The solution of unperturbed sine-Gordon equation is supplied as the initial condition. The computation is performed using OCTAVE 4.4 programming language. The present work is ended by giving the concluding remarks drawn from the computation. 

\section{Theoretical development}
\subsection{Model Hamiltonian}
The starting point of the present work is to write microscopic BCS Hamiltonian which is the total Hamiltonian of the system comprising the free Hamiltonian $(H_\text{free})$, pairing Hamiltonian $(H_\text{pair})$ and tunneling Hamiltonian $(H_\text{T})$ \cite{ambegaokar1963,sharapov2002,kim2012}, that is
\begin{equation}\label{eq1}
H=H_\text{free}+H_\text{pair}+H_\text{T},
\end{equation}
where
\begin{equation}\label{eq2}
H_\text{free}=\sum\limits_{l,i,\sigma}\int \rd^3r\creat{\text{C}}{l,i,\sigma}\biggl[\dfrac{1}{2m}(\ri\hbar\nabla+e^*\vec{A}_l^i)^2+e^*A_l^{0i}\biggr]\distroy{\text{C}}{l,i,\sigma},
\end{equation}
\begin{equation}\label{eq3}
H_\text{pair}=\sum\limits_{l,l',i,i'}\intr V_{l,l'}^{i,i'}\creat{\text{C}}{l,i,\uparrow}\creat{\text{C}}{l,i,\downarrow}\times\distroy{\text{C}}{l',i',\downarrow}\distroy{\text{C}}{l',i',\uparrow},
\end{equation}
\begin{equation}\label{eq4}
H_\text{T}=\sum\limits_{l,i,i',\sigma}\intr \biggl[T_{l,l+1}^{i,i'}\creat{\text{C}}{l,i,\sigma}\distroy{C}{l+1,i',\sigma}+T_{l+1,l}^{*i',i}\creat{C}{l+1,i',\sigma}\distroy{\text{C}}{l,i,\sigma}\biggr].
\end{equation}
Here, $\creat{\text{C}}{l,i,\sigma}(\distroy{\text{C}}{l,i,\sigma})$ is the creation(annihilation) operator for fermion with spin $\sigma=(\uparrow\text{or}\downarrow)$ for a given layer index $l$ and  band index $i$. These operators are the function of spatial coordinate $\vec{r}$ and the imaginary time $\tau=-\ri t$. $\creat{\text{C}}{l,i,\sigma}$ creates a fermion with spin $\sigma$ at the given site $(\vec{r},\tau)$ and $\distroy{C}{\sigma}$ destroy the fermion therefrom. $\creat{C}{\sigma}$ and $\distroy{C}{\sigma}$ have the dimension of inverse square root of volume (i.e. $\Omega^{-1/2}$), with $\Omega$ being the total volume of the system. $\vec{A}_l$ and $A_l^0$ are the magnetic vector potential and electric scalar potential, respectively. $e^*=2e$ and $e$ is the electronic charge and $m$ is the mass of a fermion. The operator $-i\hbar\nabla-e^*\vec{A}_l$ is called the canonical momentum operator.

Short-range or long-range phonon mediated fermions of oppsite spins form a syster of the pair of fermions. However, after the paring of such fermions, the fermionic nature is destroyed and a bosonic particle is formed. Such phonon mediated fermions having a bosonic property are called Cooper pair. $V_{l,l'}^{i,i'}$ is the coupling constant with the dimension of energy-volume (Jm$^3$). For the two-gap superconductor having s- and d-bands $i$ or $i'$ is equal to ($s,d$). $i=i'$ refers to intra-band pairing and $i\ne i'$ refers to inter-band pairing. Similarly, $l=l'$ refers to intra-layer and $l\ne l'$ refers to inter-layer pairing.

The first term of equation \eqref{eq4} infers that a fermion of spin $\sigma$ is destroyed in $(l+1)^\text{th}$ layer and $i^\text{th}$ band and is created in $l^\text{th}$ layer and ${i'}^{\text{th}}$ band, while the second term is the complex conjugate of the first term.  $T_{l,l+1}$ is the tunnel matrix element with the dimension of energy.
\subsection{Action functional}
According to the path-integral formalism of quantum mechanics, the action functional is defined as
\begin{equation}\label{eq05}
S=\int \rd t\intr \mathcal{L}
\end{equation}
with $\mathcal{L}$ being the Lagrangian density.
In terms of the total Hamiltonian, the action is defined as \cite{sharapov2002}
\begin{equation}\label{eq06}
S=\int_{0}^{\hbar\beta}\rd\tau\left\lbrace \left[\intr\sum\limits_{l,i,\sigma}\creat{\text{C}}{l,i,\sigma}\hbar\pardt\distroy{\text{C}}{l,i,\sigma}\right]+H-\mu N\right\rbrace .
\end{equation}
Here, $\mu$ is the chemical potential, and $N$  is the total particle number, $\beta=\dfrac{1}{k_{\textrm B}T}$, where $k_{\textrm B}$ is the Boltzmann constant and $T$ is absolute temperature. $\mu N$ is given as
\begin{equation}\label{eq07}
\mu N=\sum\limits_{l,i,\sigma}\intr\mu_{l,\sigma}^i\creat{\text{C}}{l,i,\sigma}\distroy{\text{C}}{l,i,\sigma} .
\end{equation}
Using equation \eqref{eq1}, \eqref{eq2}, \eqref{eq3}, \eqref{eq4},  \eqref{eq06} and \eqref{eq07}, we get the action functional as
\begin{equation}\label{eq08}
S=S_{\text{free}}+S_{\text{pair}}+S_T,
\end{equation}
\begin{eqnarray*}
S_{\text{free}}&=&\int_{0}^{\hbar\beta}\rd\tau\intr\sum\limits_{l,i,\sigma}\creat{\text{C}}{l,i,\sigma}\biggl(\hbar\pardt+\dfrac{1}{2m}(\ri\hbar\nabla+e^*\vec{A}_l^i)^2+e^*A_l^{0i}-\mu_{l,\sigma}^i\biggr)\distroy{\text{C}}{l,i,\sigma},\\
\nonumber
S_{\text{pair}}&=&\int_{0}^{\hbar\beta}\rd\tau\intr\sum\limits_{l,l',i,i'}V_{l,l'}^{i,i'}\creat{\text{C}}{l,i,\uparrow}\creat{\text{C}}{l,i,\downarrow}\times\distroy{\text{C}}{l',i',\downarrow}\distroy{\text{C}}{l',i',\uparrow},\\
\nonumber
S_{T}&=&\intau\intr\sum\limits_{l,i,i',\sigma}\biggl[T_{l,l+1}^{i,i'}\creat{\text{C}}{l,i,\sigma}\distroy{C}{l+1,i',\sigma}+T_{l+1,l}^{*i',i}\creat{C}{l+1,i',\sigma}\distroy{\text{C}}{l,i,\sigma}\biggr].
\end{eqnarray*}
Now, the quantum mechanical partition function of the system can be written as
\begin{equation}\label{eq09}
Z=\int\mathcal{D}[\text{C}^\dagger,\text{C}]\exp\left(-\dfrac{S_\text{free}}{\hbar}-\dfrac{S_\text{pair}}{\hbar}-\dfrac{S_T}{\hbar}\right).
\end{equation}
Here, $\text{C}$ is a column vector with elements $\distroy{\text{C}}{l,i,\sigma}$, and $\text{C}^\dagger$ is a row vector with elements $\creat{\text{C}}{l,i,\sigma}$ while $\int\mathcal{D}[\text{C}^\dagger,\text{C}]$ represents the product of all integrals over the elements of $\text{C}^\dagger$ and $\text{C}$.

\subsection{Hubbard-Stratonovich transformation}
The action functional associated with the pair Hamiltonian is in quartic form of four fermionic fields.  Since $\vec{A}_l^i$ and $A_l^{0i}$ are invariant under gauge transformation, the partition function of \eqref{eq09} can be rewritten as follows: 
\begin{multline}\label{eq25}
Z=\int\mathcal{D}[\text{C}^\dagger,\text{C}]\exp\biggl\{-\dfrac{1}{\hbar}\intau\intr\biggl[\sum\limits_{l,i,\sigma}\creat{\text{C}}{l,i,\sigma}\times\left(\hbar\pardt-\dfrac{\hbar^2}{2m}\nabla^2-\mu_{l,\sigma}^i\right)\distroy{\text{C}}{l,i,\sigma}\\
-\sum\limits_{l,l',i,i'}V_{l,l'}^{i,i'}\creat{\text{C}}{l,i,\uparrow}\creat{\text{C}}{l,i,\downarrow}\distroy{\text{C}}{l',i',\downarrow}\distroy{\text{C}}{l',i',\uparrow}\\
+\sum\limits_{l,i,i',\sigma}\biggl(T_{l,l+1}^{i,i'}\creat{\text{C}}{l,i,\sigma}\distroy{C}{l+1,i',\sigma}+T_{l+1,l}^{*i',i}\creat{C}{l+l,i',\sigma}\distroy{\text{C}}{l,i,\sigma}\biggr)\biggr]\biggr\}.
\end{multline}
Under the application of Hubbard-Stratonovich transformation, the quartic term of the pairing interaction can be reduced to quadratic, and the partition function becomes
\begin{eqnarray}
Z&=&\int\mathcal{D}[\bar{\Delta},\Delta]\int\mathcal{D}[\text{C}^\dagger,C]\exp\biggl\{-\dfrac{1}{\hbar}\intau\intr\\ \nonumber
&\times&\biggl[\sum\limits_{l,i,\sigma}\creat{\text{C}}{l,i,\sigma}\left(\hbar\pardt-\dfrac{\hbar^2}{2m}\nabla^2-\mu_{l,\sigma}^i\right)\distroy{\text{C}}{l,i,\sigma}\\\nonumber
&+&\sum\limits_{l,l',i,i'}\biggl(\bar{\Delta}_{l,i}(V^{-1})_{l,l'}^{i,i'}\Delta_{l',i'}+\bar{\Delta}_{l,i}\distroy{C}{l,i,\downarrow}\distroy{C}{l,i,\uparrow}+\Delta_{l,i}\creat{\text{C}}{l,i,\uparrow}\creat{\text{C}}{l,i,\downarrow}\biggr)\\\nonumber
&+&\sum\limits_{l,i,i',\sigma}\biggl(T_{l,l+1}^{i,i'}\creat{\text{C}}{l,i,\sigma}\distroy{C}{l+1,i',\sigma}+T_{l+1,l}^{*i',i}\creat{C}{l+1,i',\sigma}\distroy{\text{C}}{l,i,\sigma}\biggr)\biggr]\biggr\},
\label{eq27}
\end{eqnarray}
where, $\bar{\Delta}(\Delta)$ is the new fields which are bosonic in nature. $\bar{\Delta}$ is a row vector containing the elements $\bar{\Delta}_{l,i}(\vec{r},\tau)$ and $\Delta$ is a column vector containing the elements $\Delta_{l,i}(\vec{r},\tau)$. 
Applying  special techniques of path integral formalism in this partition function and performing some matrix manipulation we could obtain the Lagrangian density as follows:
\begin{eqnarray}
\mathcal{L}&=&\sum\limits_{l,l',i,i'}\Delta_{0li}^*(V^{-1})_{ll'}^{ii'}\Delta_{0l'i'}\re^{-\ri(\theta_{li}-\theta_{l'i'})}
+\sum\limits_{l,i}\left(\dfrac{\hbar^2 N(0)}{4}\right)\left(\pard{\theta_{li}}{\tau}+\dfrac{e^*A_l^{0i}}{\hbar}\right)^2 \nonumber \\ 
&+&\sum\limits_{l,i}\left(\dfrac{\hbar^2 N(0)\mu_l^i}{6m}\right)\left(\nabla\theta_{li}-\dfrac{e^*\vec{A}_l^i}{\hbar}\right)^2
-\sum\limits_{l,i,i'}\bigg[\dfrac{2 T_{l,l+1}^{ii'}T_{l+1,l}^{i'i} N(0)}{\Delta_{0l+1,i'}^2-\Delta_{0li}^2}\ln\left(\dfrac{\Delta_{0l+1,i'}}{\Delta_{0li}}\right) \nonumber\\
&\times&\Delta_{0li}\Delta_{0l+1,i'}\cos(\theta_{l+1,i'}-\theta_{li})+2 N(0)\hbar\omega_D\zeta_l^i\delta_{ii'}+ N(0)\hbar^2\omega_D^2\delta_{ii'}\biggr].
\label{lag:01}
\end{eqnarray} 
At low temperature, the chemical potential $\mu_l^i$ is equal to the Fermi energy, i.e., $\mu_l^i=\epsilon_F$ and $\zeta_l^i=0$ since $\mu_\uparrow=\mu_\downarrow$. 
We also have, 
\[ N(0)=\dfrac{3n}{4\epsilon_F}=\dfrac{3}{4}\dfrac{\textbf{k}_\text{F}^2}{3\piup^2}\dfrac{2m}{\hbar^2\textbf{k}_\text{F}^2}=\dfrac{m\textbf{k}_\text{F}}{2\piup^2\hbar^2}. \]
The effective Lagrangian is given by
\begin{eqnarray}
&&\mathcal{L}_\text{eff}=\sum\limits_{l,i}\dfrac{\varepsilon_0}{2\lambda_{\text{TF}}^2}\left(\dfrac{\hbar}{e^*}\pard{\theta_{li}}{\tau}+A_l^{0i}\right)^2
+\sum\limits_{l,i}\dfrac{\varepsilon_0 c^2}{2\lambda_\text{L}^2}\left(\dfrac{\hbar}{e^*}\nabla\theta_{li}-\vec{A}_l^i\right)^2\nonumber \\
&&+\sum\limits_{l,l',i,i'}\Delta_{0li}^*(V^{-1})_{ll'}^{ii'}\Delta_{0l'i'}\re^{-\ri(\theta_{li}-\theta_{l'i'})}-\sum\limits_{l,i,i'}\bigg[\dfrac{2 T_{l,l+1}^{ii'}T_{l+1,l}^{i'i} N(0)}{\Delta_{0l+1,i'}^2-\Delta_{0li}^2}\ln\left(\dfrac{\Delta_{0l+1,i'}}{\Delta_{0li}}\right)\nonumber\\
&&\times\Delta_{0li}\Delta_{0l+1,i'}\cos(\theta_{l+1,i'}-\theta_{li})+ N(0)\hbar^2\omega_D^2\delta_{ii'}\biggr]
+\sum_{l,i}\left[\dfrac{\varepsilon_{rb}\varepsilon_0}{2}(E_l^i)^2+\dfrac{\varepsilon_{rb}\varepsilon_0c^2}{2}(B_l^i)^2\right],
\label{eq10.22}
\end{eqnarray} 
where, $n$ is the concentration of electronic charge, $\textbf{k}_\text{F}$ is the Fermi wave vector, $\lambda_{\text{TF}}=\sqrt{\dfrac{\varepsilon_0\piup^2\hbar^2}{e^2m\textbf{k}_\text{F}}}$ is the Thomas-Fermi charge screening length and $\lambda_\text{L}=\sqrt{\dfrac{\varepsilon_0mc^2}{ne^2}}$ is the London penetration depth, $\vec{E}_l^i$ and $\vec{B}_l^i$ are electric and magnetic fields at layer $l$ and $ i $ band.
\subsubsection{Application to the long Josephson junction}
Consider the stack of long Josephson junction with the lenght along $x$-direction and junction system along $z$-direction. External magnetic fields are applied along the $y$-direction, which introduces a homogeneous phase difference along the $x$-direction. The system is assumed to be  uniform along the $y$-direction and the problem becomes two dimensional. The electric field is along $z$-direction. Now, the Lagrangian density in two-dimensional system becomes, as follows:
\begin{eqnarray}
\mathcal{L}_\text{eff}&=&\dfrac{\varepsilon_0d}{2\lambda_{\text{TF}}^2}\sum\limits_{l,i}\left(\dfrac{\hbar}{e^*}\pard{\theta_l^i}{\tau}+A_l^{0i}\right)^2
+\dfrac{\varepsilon_0 c^2d}{2\lambda_\text{L}^2}\sum\limits_{l,i}\left(\dfrac{\hbar}{e^*}\pard{\theta_l^i}{x}-A_l^{xi}\right)^2
+\sum\limits_{l,i,i'}\dfrac{\hbar}{e^*}J_{ll}^{ii'}\cos(\theta_{li}-\theta_{li'})\nonumber\\
&-&\sum\limits_{l,i,i'}\bigg[\dfrac{\hbar}{e^*}j_{l,l+1}^{ii'}\cos(\theta_{l+1,i'}-\theta_{li})+ N(0)d\hbar^2\omega_D^2\delta_{ii'}\biggr]\nonumber\\
&+&\sum_{l,i,i'}\left[\dfrac{\varepsilon_{rb}\varepsilon_0b}{2}(E_{l,l+1}^{zii'})^2+\dfrac{\varepsilon_{rb}\varepsilon_0c^2b}{2}
(B_{l,l+1}^{yii'})^2\right],
\label{eq11.01}
\end{eqnarray} 
where $d$ is the thickness of the superconducting layer and $b$ is the thickness of the junction material. $\varepsilon_{rb}$ is the dielectric constant of the junction material. The inter-band Josephson coupling constant is
\begin{equation}\label{eq11.02}
J_{ll}^{ii'}=\dfrac{e^*d}{\hbar}\Delta_{0li}^*(V^{-1})_{ll}^{ii'}\Delta_{0li'}
\end{equation}
and Josephson tunneling coupling constant 
\begin{equation}\label{eq11.03}
j_{l,l+1}^{ii'}=\dfrac{e^*d}{\hbar}\dfrac{2 T_{l,l+1}^{ii'}T_{l+1,l}^{i'i} N(0)}{\Delta_{0l+1,i'}^2-\Delta_{0li}^2}\ln\left(\dfrac{\Delta_{0l+1,i'}}{\Delta_{0li}}\right)\Delta_{0li}\Delta_{0l+1,i'}\\+\dfrac{e^*d}{\hbar}\Delta_{0li}^*(V^{-1})_{l,l+1}^{ii'}\Delta_{0,l+1,i'}.
\end{equation}
The $z$-component of the electric field in between  $l^{\text{th}}$ and $(l+l)^\text{th}$ layer is 
\begin{equation}\label{11.04}
E_{l,l+1}^{zii'}=-\pard{A_{l,l+1}^{zii'}}{t}-\dfrac{1}{b}\left(A_{l+1}^{0i'}-A_l^{0i}\right)
\end{equation}
and the $y$-component of the magnetic field in between  $l^{\text{th}}$  and $(l+l)^\text{th}$ layer is
\begin{equation}\label{eq11.05}
B_{l,l+1}^{yii'}=\dfrac{1}{b}\left(A_{l+1}^{xi'}-A_l^{xi}\right)-\pard{A_{l,l+1}^{zii'}}{x}
\end{equation}
with
\begin{equation}\label{eq11.06}
A_{l,l+1}^{zii'}=\dfrac{1}{b}\int_{-b/2}^{+b/2}A^z(z)dz.
\end{equation}
We can introduce the gauge invariant phase difference $\varphi_{l,l+1}^{ii'}$ as 
\begin{equation}\label{eq11.07}
\varphi_{l,l+1}^{ii'}=\theta_{l+1}^{i'}-\theta_l^i-\dfrac{be^*}{\hbar}A_{l,l+1}^{zii'}.
\end{equation}
Then, we can have $\cos\left(\theta_{l+1}^{i'}-\theta_l^i\right)=\cos\varphi_{l,l+1}^{ii'}$ 
and $\theta_{l}^{i'}-\theta_l^i=\chi_{ll}^{ii'}$ is the intra-layer inter-band phase difference. 
Hence, the equation \eqref{eq11.01} 
\begin{eqnarray}
\mathcal{L}_\text{eff}&=&\dfrac{\varepsilon_0d}{2\lambda_{\text{TF}}^2}\sum\limits_{l,i}\left(\dfrac{\hbar}{e^*}\pard{\theta_l^i}
{\tau}+A_l^{0i}\right)^2+\dfrac{\varepsilon_0 c^2d}{2\lambda_\text{L}^2}\sum\limits_{l,i}\left(\dfrac{\hbar}{e^*}\pard{\theta_l^i}{x}
-A_l^{xi}\right)^2\nonumber\\
&+&\sum\limits_{l,i,i'}\dfrac{\hbar}{e^*}J_{ll}^{ii'}\cos\chi_{ll}^{ii'}-\sum\limits_{l,i,i'}\bigg[\dfrac{\hbar}{e^*}j_{l,l+1}^{ii'}
\cos\varphi_{l,l+1}^{ii'}+ N(0)d\hbar^2\omega_D^2\delta_{ii'}\biggr]
+\sum_{l,i,i'}\biggl[\dfrac{\varepsilon_{rb}\varepsilon_0b}{2}\nonumber\\
&\times&\left(\pard{A_{l,l+1}^{zii'}}{t}+\dfrac{1}{b}\left(A_{l+1}^{0i'}-A_l^{0i}\right)\right)^2+\dfrac{\varepsilon_{rb}\varepsilon_0c^2b}{2}\left(\dfrac{1}{b}\left(A_{l+1}^{xi'}-A_l^{xi}\right)-\pard{A_{l,l+1}^{zii'}}{x}\right)^2\biggr].
\label{eq11.08}
\end{eqnarray} 
The Lagrangian density equation \eqref{eq11.08} can be minimized using the Euler-Lagrange equation. Applying the Euler-Lagrange equation with respect to $A_k^{0j}$,  $A_{k+1}^{0j'}$, $A_k^{xj}$,  $A_{k+1}^{xj'}$, $A_{k,k+1}^{zjj'}$ and $\theta_k^j$ with $k$ as a new layer index and $j,j'$ as a new band index, we get the generalized equation for the phase dynamics applicable for homogeneous superconducting layers
\begin{eqnarray}
&&\dfrac{\varepsilon_{rb}bd}{\lambda_F^2}\pard{^2\varphi_{k,k+1}^{jj'}}{\bar{t}^2}+\varepsilon_{rb}^2\sum_{i}\pard{^2}{\bar{t}^2}\biggl
(2\varphi_{k,k+1}^{ij'}-\varphi_{k-1,k}^{ij}-\varphi_{k+1,k+2}^{j'i}\biggr)-\dfrac{\varepsilon_{rb}bd}
{\lambda_{F}^2}\pard{^2\varphi_{k,k+1}^{jj'}}{\bar{x}^2}
\nonumber\\
&&-\varepsilon_{rb}^2\sum_i\pard{^2}{\bar{x}^2}\left(2\varphi_{k,k+1}^{ij'}-\varphi_{k-1,k}^{ij}-\varphi_{k+1,k+2}^{j'i}\right)
+\dfrac{b^2d^2}{\lambda_{L}^2\lambda_{\text{TF}}^2J_0}j_{k,k+1}^{jj'}\sin\varphi_{k,k+1}^{jj'}\nonumber\\
&&+\dfrac{\varepsilon_{rb}bd}{J_0}\left(\dfrac{1}{\lambda_{\text{TF}}^2}+\dfrac{1}{\lambda_\text{L}^2}\right)\sum_i\biggl(2j_{k,k+1}^{ij'}\sin\varphi_{k,k+1}^{ij'}-j_{k-1,k}^{ij}\sin\varphi_{k-1,k}^{ij}-j_{k+1,k+2}^{j'i}\sin\varphi_{k+1,k+2}^{j'i}\biggr)\nonumber\\
&&+\dfrac{2\varepsilon_{rb}^2}{J_0}\sum_{i,i'}\biggl(2j_{k,k+1}^{i'j'}\sin\varphi_{k,k+1}^{i'j'}-j_{k-1,k}^{i'i}\sin\varphi_{k-1,k}^{i'i}-j_{k+1,k+2}^{j'i'}\sin\varphi_{k+1,k+2}^{j'i'}\biggr)\nonumber\\
&&+\dfrac{\varepsilon_{rb}^2}{J_0}\sum_{i,i'}\biggl(2j_{k-1,k}^{i'j}\sin\varphi_{k-1,k}^{i'j}-j_{k-2,k-1}^{i'i}\sin\varphi_{k-2,k-1}^{i'i}-j_{k,k+1}^{ji'}\sin\varphi_{k,k+1}^{ji'}\biggr)\nonumber\\
&&+\dfrac{\varepsilon_{rb}^2}{J_0}\sum_{i,i'}\biggl(2j_{k+1,k+2}^{i'i}\sin\varphi_{k+1,k+2}^{i'i}-j_{k,k+1}^{ij'}\sin\varphi_{k,k+1}^{ij'}-j_{k+2,k+3}^{j'i}\sin\varphi_{k+2,k+3}^{ii'}\biggr)=0.
\label{eq11.36}
\end{eqnarray}
\subsubsection{Coupled long Josephson junction system}
\begin{figure}[!t]
	\centering
	\includegraphics[width=3in]{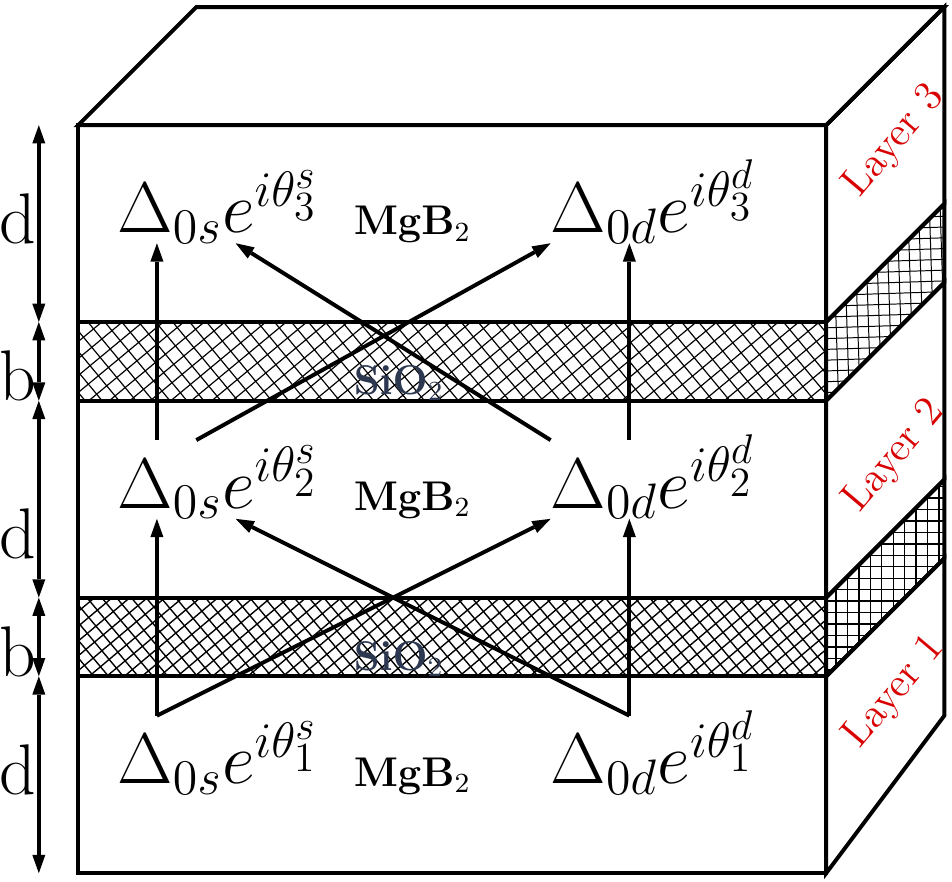}
	\caption{(Colour online) A typical coupled LJJ}
	\label{doublejunction}
\end{figure}

In the
coupled long Josephson junction,  as shown in figure~\ref{doublejunction}, there are eight
channels for Cooper pair tunneling. The equations of phase dynamics can be obtained from the generalized equation \eqref{eq11.36} as
\begin{equation}\label{dble:5}
\pard{^2\varphi}{\bar{t}^2}-\pard{^2\varphi}{\bar{x}^2}+\mathcal{M}_{0d}^{-1}\mathcal{M}_{Fd}(\bar{j}\sin\varphi)=0
\end{equation} 
with 
\[\varphi=\begin{pmatrix}
\varphi_{12}\\\varphi_{23}
\end{pmatrix},\quad\varphi_{12}=\begin{pmatrix}
\varphi_{12}^{ss}\\\varphi_{12}^{sd}\\\varphi_{12}^{ds}\\\varphi_{12}^{dd}
\end{pmatrix},\quad\varphi_{23}=\begin{pmatrix}
\varphi_{23}^{ss}\\\varphi_{23}^{sd}\\\varphi_{23}^{ds}\\\varphi_{23}^{dd}
\end{pmatrix},\]
\[\bar{j}=\begin{pmatrix}
\bar{j}_{12} & 0\\ 0 &\bar{j}_{23}
\end{pmatrix},\quad\bar{j}_{12}=\begin{pmatrix}
\bar{j}_{12}^{ss}& 0 & 0 & 0\\0 & \bar{j}_{12}^{sd} & 0 & 0\\0 & 0 & \bar{j}_{12}^{ds}&0\\0 & 0 & 0&\bar{j}_{12}^{dd}
\end{pmatrix},\]\[\quad\bar{j}_{23}=\begin{pmatrix}
\bar{j}_{23}^{ss} & 0 & 0 & 0\\0 & \bar{j}_{23}^{sd}& 0 & 0\\0 & 0 & \bar{j}_{23}^{ds}& 0\\0 & 0 &0 &\bar{j}_{23}^{dd}
\end{pmatrix}, \quad\bar{j}_{12}^{ss}=\dfrac{j_{12}^{ss}}{J_0},\quad\text{and so on.}\]
\[\mathcal{M}_0d=\begin{pmatrix}
\mathcal{M}_0&\mathcal{M}_1\\
\mathcal{M}_1&\mathcal{M}_0
\end{pmatrix},\quad\mathcal{M}_0=\begin{pmatrix}
\alpha_0 & 0 & 2\varepsilon_{rb}^2 & 0\\
0 & \alpha_0 & 0 & 2\varepsilon_{rb}^2\\
2\varepsilon_{rb}^2 & 0 &\alpha_0 & 0\\
0 & 2\varepsilon_{rb}^2 & 0 & \alpha_0
\end{pmatrix},\]
\[\mathcal{M}_1=\begin{pmatrix}
-\varepsilon_{rb}^2 & -\varepsilon_{rb}^2 & 0 & 0\\
-\varepsilon_{rb}^2 & -\varepsilon_{rb}^2 & 0 & 0\\
0 & 0 & -\varepsilon_{rb}^2 & -\varepsilon_{rb}^2\\
0 & 0 &-\varepsilon_{rb}^2 & -\varepsilon_{rb}^2 
\end{pmatrix},\]
\[\mathcal{M}_{Fd}=\begin{pmatrix}
\mathcal{M}_{F1} & \mathcal{M}_{F2}\\
-\mathcal{M}_{F2} & \mathcal{M}_{F1}
\end{pmatrix},\]\[\mathcal{M}_{F1}=\begin{pmatrix}
\beta_0 & -\varepsilon_{rb}^2 & \beta_1 & 0\\
-\varepsilon_{rb}^2 & \beta_0 & 0 & \beta_1\\
\beta_1 & 0 & \beta_0 & -\varepsilon_{rb}^2\\
0 & \beta_1 & -\varepsilon_{rb}^2 & \beta_0 
\end{pmatrix},\]
\[\mathcal{M}_{F2}=\begin{pmatrix}
-\beta_2 & -\beta_2 & 2\varepsilon_{rb}^2 & 2\varepsilon_{rb}^2\\
-\beta_2 & -\beta_2 & 2\varepsilon_{rb}^2 & 2\varepsilon_{rb}^2\\
2\varepsilon_{rb}^2 & 2\varepsilon_{rb}^2&-\beta_2 & -\beta_2\\
2\varepsilon_{rb}^2 & 2\varepsilon_{rb}^2&-\beta_2 & -\beta_2
\end{pmatrix},\]
\begin{eqnarray*}
\alpha_0&=&\dfrac{\varepsilon_{rb}bd}{\lambda_{F}^2}+2\varepsilon_{rb}^{2}\,,\nonumber\\
\beta_0&=&\dfrac{b^2d^2}{\lambda_{L}^2\lambda_{F}^2}+2\varepsilon_{rb}bd\left(\dfrac{1}{\lambda_{\text{TF}}^2}+\dfrac{1}{\lambda_{L}^2}\right)
+2\varepsilon_{rb}^2\,,\nonumber\\
\beta_1&=&\varepsilon_{rb}bd\left(\dfrac{1}{\lambda_{\text{TF}}^2}+\dfrac{1}{\lambda_{L}^2}\right)+3\varepsilon_{rb}^2\,,\nonumber\\
\beta_2&=&\varepsilon_{rb}bd\left(\dfrac{1}{\lambda_{\text{TF}}^2}+\dfrac{1}{\lambda_{L}^2}\right), \quad \text{and}\quad J_0=\dfrac{\varepsilon_0dc^2\hbar}{\lambda_{\text{TF}}^2\lambda_\text{L}^2e^*}.
\end{eqnarray*}

In order to study the plasmon mode, the equation \eqref{dble:5} can be linealized as
\begin{equation}\label{dble:5a}
\pard{^2\varphi}{\bar{t}^2}-\pard{^2\varphi}{\bar{x}^2}+\mathcal{M}_{0d}^{-1}\mathcal{M}_{Fd}(\bar{j}\varphi)=0
\end{equation} 
for small phase differences $\varphi$. The equation \eqref{dble:5a} has the solution $\varphi=\varphi_0\exp[\ri(\bar{\omega}\bar{t}\pm\bar{k}\bar{x})]$ with dispersion relation
\begin{equation}\label{disp}
\bar{\omega}=\sqrt{\mathcal{M}_{0d}^{-1}\mathcal{M}_{Fd}\bar{j}+\bar{k}^2}\,,
\end{equation}
where $\bar{\omega}$ and $\bar{k}$ are the normalized frequency and wave vector, respectively.

\section{Numerical computation and analysis} 
\begin{figure}[!t]
	\begin{subfigure}[t]{0.49\columnwidth}
		\centering
		\includegraphics[width=\textwidth]{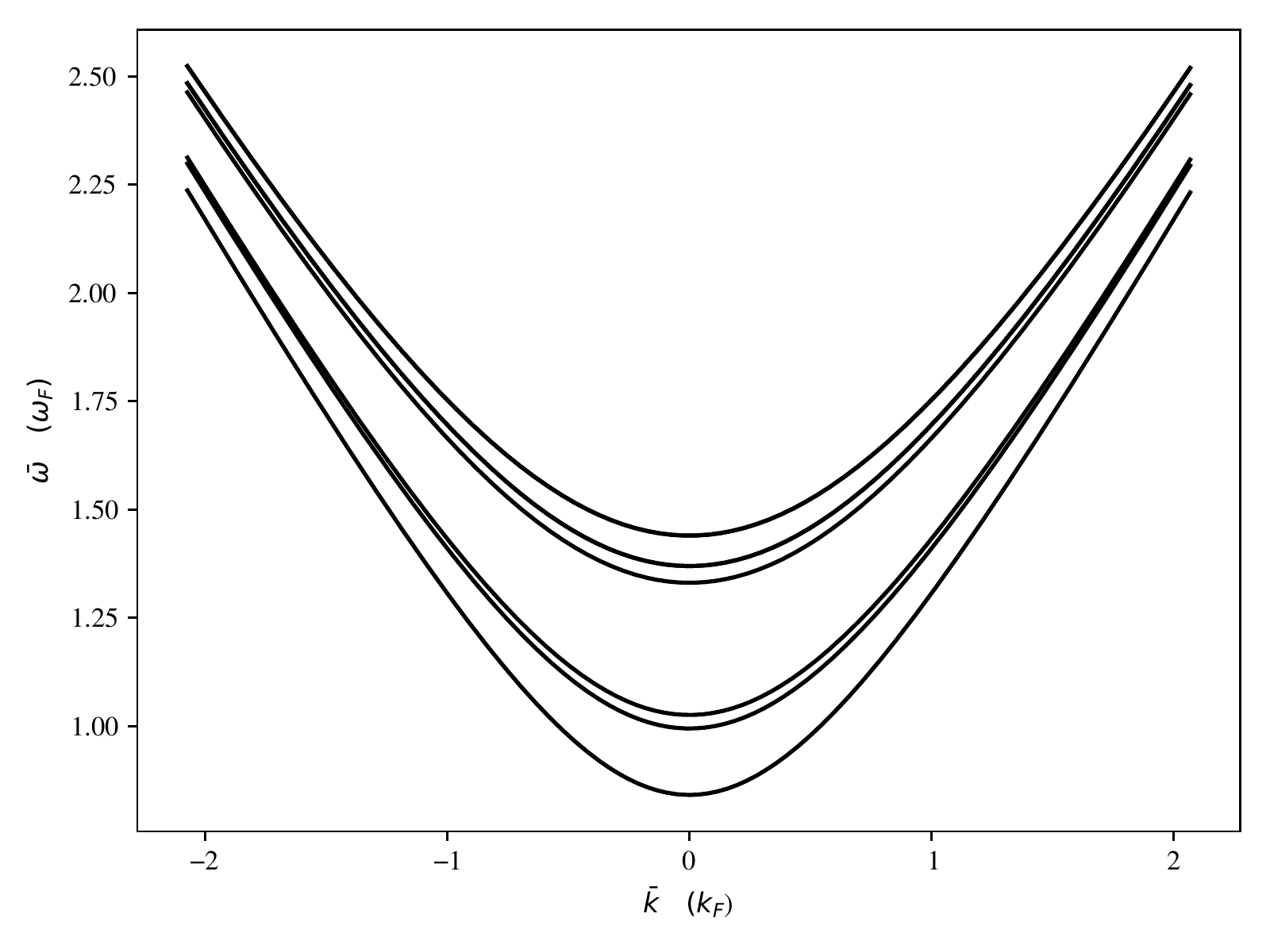}
		\caption{} \label{plasma:a}
	\end{subfigure}
	\begin{subfigure}[t]{0.49\columnwidth}
		\centering
		\includegraphics[width=\textwidth]{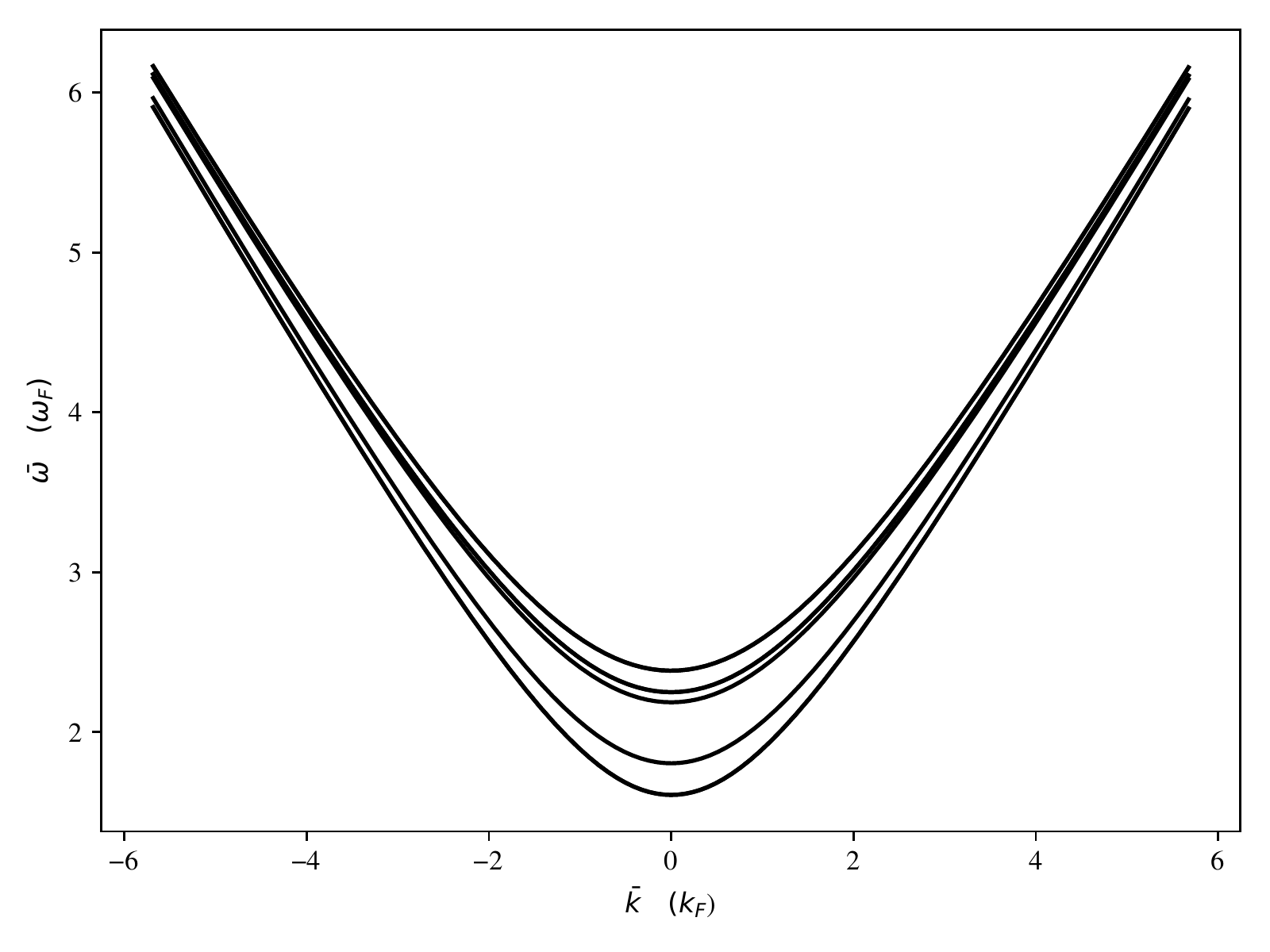}
		\caption{} \label{plasma:b}
	\end{subfigure}
	\begin{subfigure}[t]{0.49\columnwidth}
		\centering
	\includegraphics[width=\textwidth]{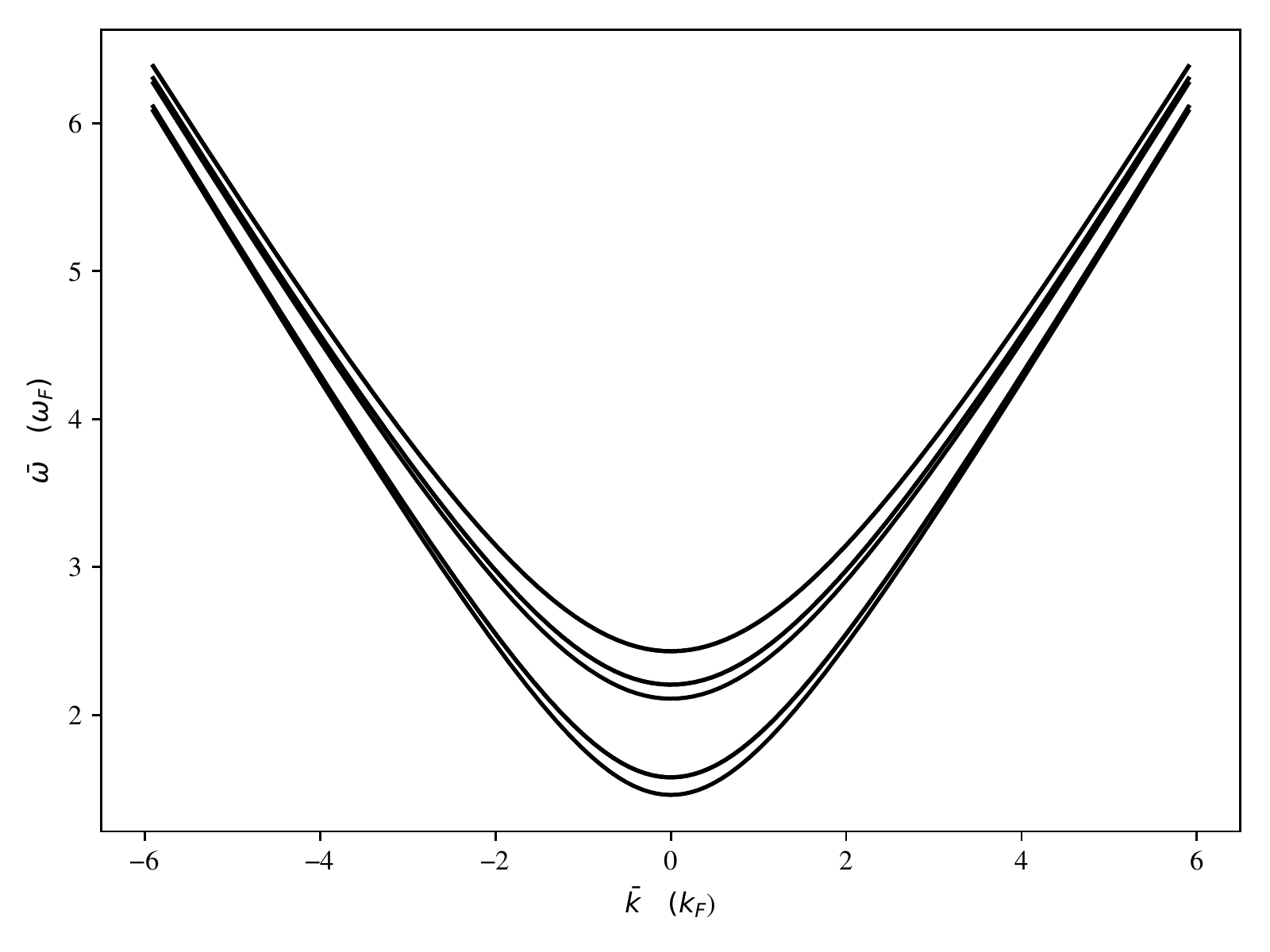}
		\caption{} \label{plasma:c}
	\end{subfigure}
	\begin{subfigure}[t]{0.49\columnwidth}
		\centering
		\includegraphics[width=\textwidth]{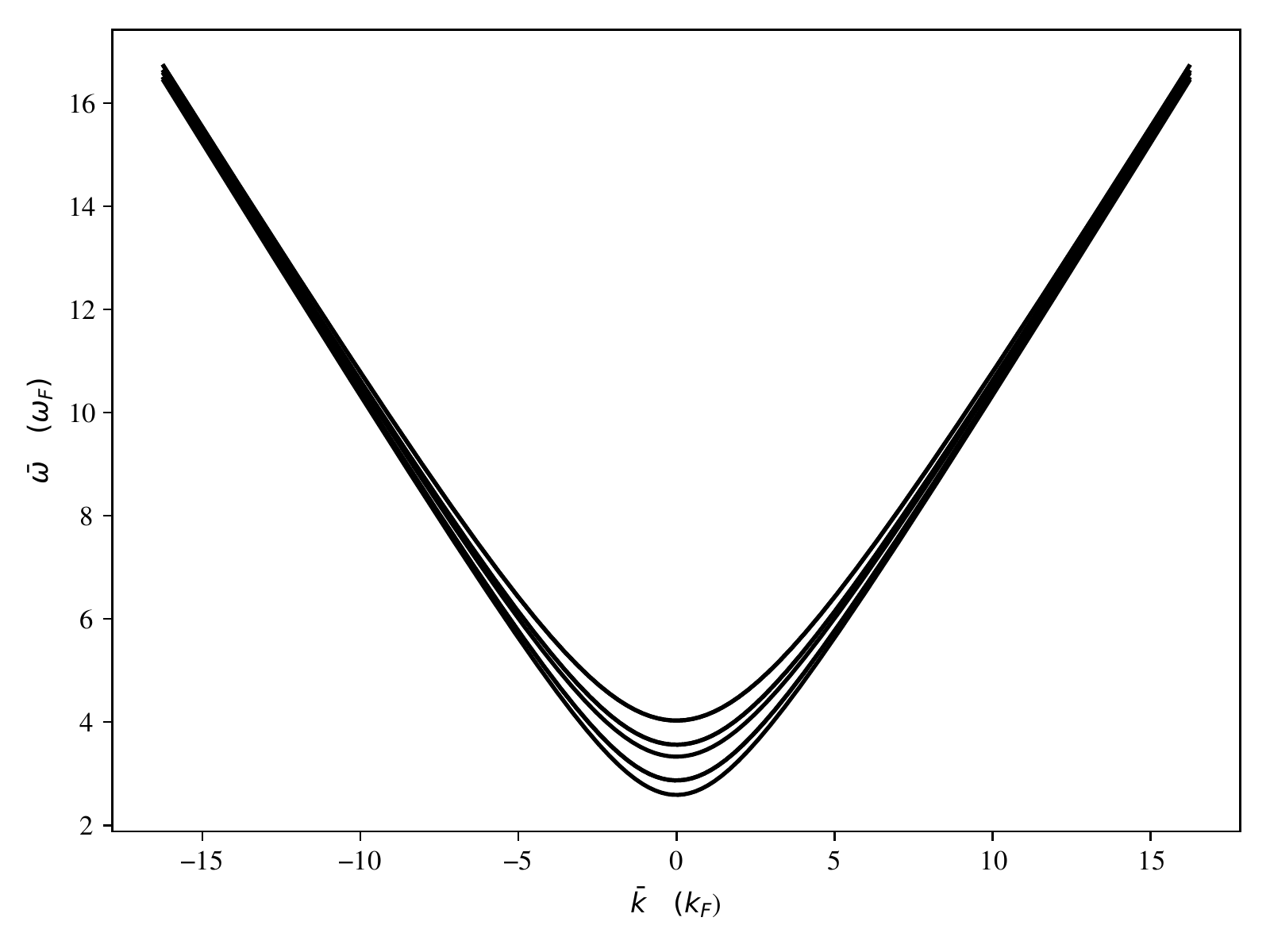}
		\caption{} \label{plasma:d}
	\end{subfigure}
\caption{Dispersion relation for (\subref{plasma:a}) $b=3$ {\AA}, $d=6$ {\AA}, $V= 0.5$ V, (\subref{plasma:b}) $b=3$ {\AA}, $d=6$ {\AA}, $V= 1$ V, (\subref{plasma:c}) $b=6$ {\AA}, $d=9$ {\AA}, $V= 0.5$ V and (\subref{plasma:d}) $b=6$ {\AA}, $d=9$ {\AA}, $V= 1$ V. The plots show that the plasma wave is at the excited state as the junction and layer thicknesses are increased. The excitation also depends on the applied voltages.}\label{plasma}
\end{figure}
In order to perform the numerical computation, the equation \eqref{dble:5} is discretized using the finite difference approximation. For this purpose, a uniform mesh in space and time is introduced with spacing $\delta x$ and $\delta t$, respectively. At each point $(\bar{x}_i,\bar{t}_n)$
\begin{gather}
\bar{x}_i=-L_x+i\delta x, \text{ with } i=0,\ldots,N_x,\label{eq4.05}\\
\bar{t}_n=n\delta t, \text{ with } n=0,\ldots,N_t\,,\label{eq4.06}
\end{gather}
where $N_x$ and $N_t$ are the total number of the points in space and time, respectively. The sine-Gordon equation is approximated by the second-order finite differences as
\begin{gather}
\pard{^2\varphi(\bar{x}_i,\bar{t}_n)}{\bar{t}^2}\approx\dfrac{\varphi_i^{n+1}-2\varphi_i^n+\varphi_i^{n-1}}{\delta t^2},\label{eq4.07}\\
\pard{^2\varphi(\bar{x}_i,\bar{t}_n)}{\bar{x}^2}\approx\dfrac{\varphi_{i+1}^n-2\varphi_i^n+\varphi_{i-1}^n}{\delta x^2}\,,\label{eq4.08}
\end{gather}
where $\varphi_i^n$ is the numerical approximation of the exact solution at $(\bar{x}_i,\bar{t}_n)$. Applying this approximation, the perturbed sine-Gordon equation reads
\begin{equation}
\varphi_i^{n+1}=-\varphi_i^{n-1}+2\varphi_i^n+\dfrac{\delta t^2}{\delta x^2}\left(\varphi_{i+1}^n-2\varphi_i^n+\varphi_{i-1}^n\right)-\delta t^2\mathcal{M}_{0d}^{-1}\mathcal{M}_{Fd}(\bar{j}\sin\varphi_i^n).
\label{eq4.09}
\end{equation}
\begin{figure}[!t]
\centering
	\includegraphics[clip,width=0.98\textwidth,angle=0]{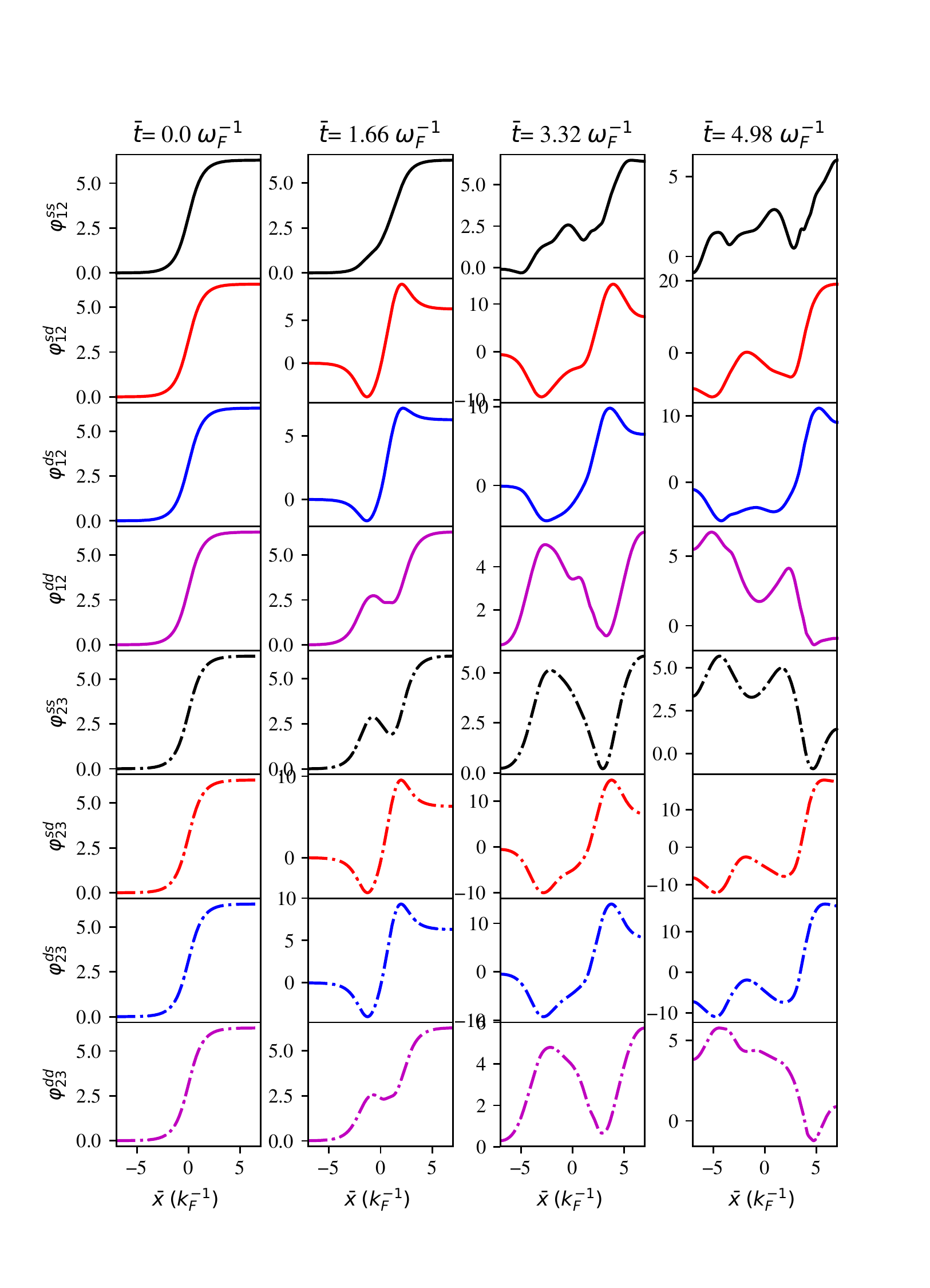}
	\caption{(Colour online) Spatial-temporal variation of phase differences in  various channels in the coupled LJJ with junction thickness of 3 {\AA} and layer thickness of 6 {\AA} under the application of bias voltage of 0.5 V.}\label{phase:01}
\end{figure}
Providing the initial conditions  to the junction system means supplying the initial information to the system at the starting time. In the present problem, the initial information is the kink (or anti-kink) solution of unperturbed sine-Gordon equation which can be generated by the appropriate electronic device which produces it as the trigger signal~\cite{krasnov1997}. The solution of unperturbed sine-Gordon equation is 
\begin{equation}\label{eq4.02}
\varphi(\bar{x},\bar{t})=4\tan^{-1}\left[\exp\left(\sigma\dfrac{\bar{x}-u\bar{t}-\bar{x}_0}{\sqrt{1-u^2}}\right)\right],
\end{equation}
\begin{figure}[!t]
\centering
	\includegraphics[clip,width=0.97\textwidth,angle=0]{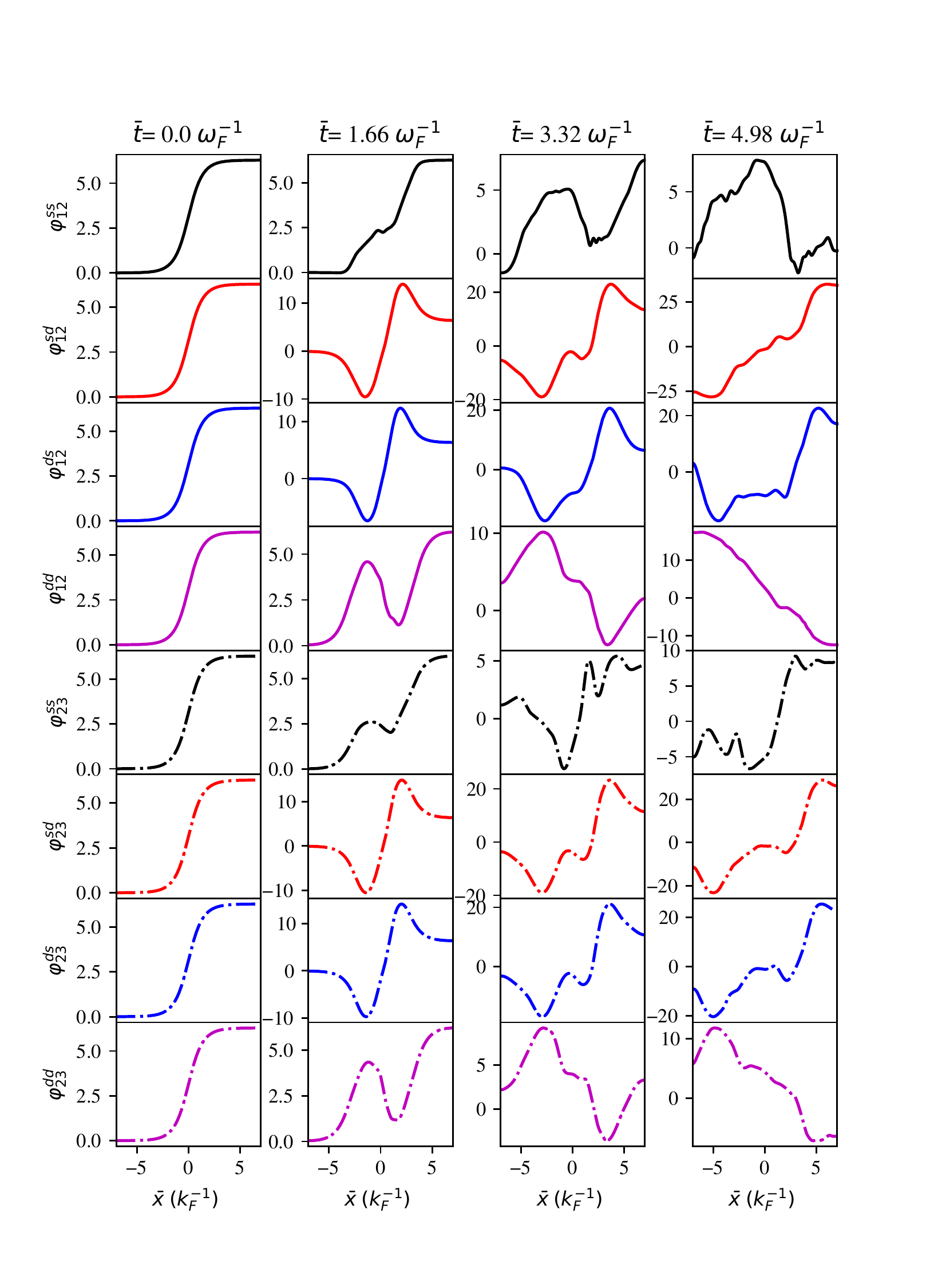}
	\caption{(Colour online) Spatial-temporal variation of phase differences in  various channels in the coupled LJJ with junction thickness of 3 {\AA} and layer thickness of 6 {\AA} under the application of bias voltage of 1 V.}\label{phase:02}
\end{figure}
where $u$ is the normalized speed of the kink ($\sigma=+1$) or anti-kink ($\sigma =-1$) and $\bar{x}_0$ is its  initial position. Hence, the initial condition for all channels of the junction system is 
\begin{equation}\label{eq4.03}
\varphi(\bar{x},0)=4\tan^{-1}\left[\exp\left(\sigma\dfrac{\bar{x}-\bar{x}_0}{\sqrt{1-u^2}}\right)\right]
\end{equation}
and
\begin{equation}\label{eq4.04}
\left.\pard{\varphi}{\bar{t}}\right|_{\bar{t}=0}=-2\sigma\dfrac{u}{\sqrt{1-u^2}}\sech\left(\sigma\dfrac{\bar{x}-\bar{x}_0}{\sqrt{1-u^2}}\right).
\end{equation}
The initial condition is  approximated as
\begin{gather}
\varphi(\bar{x}_i,0)=\varphi_i^0=4\tan^{-1}\left[\exp\left(\sigma\dfrac{\bar{x}_i-\bar{x}_0}{\sqrt{1-u^2}}\right)\right],\label{eq4.11}\end{gather}
\begin{gather}
\left.\pard{\varphi}{\bar{t}}\right|_{\bar{t}=0}\approx\dfrac{\varphi_i^1-\varphi_i^{-1}}{2\delta t}=-\dfrac{2\sigma u}{\sqrt{1-u^2}}\sech\left[\sigma\dfrac{\bar{x}_i-\bar{x}_0}{\sqrt{1-u^2}}\right].
\label{eq4.13}
\end{gather}

\begin{figure}[!t]
\centering
	\includegraphics[clip,width=0.98\textwidth,angle=0]{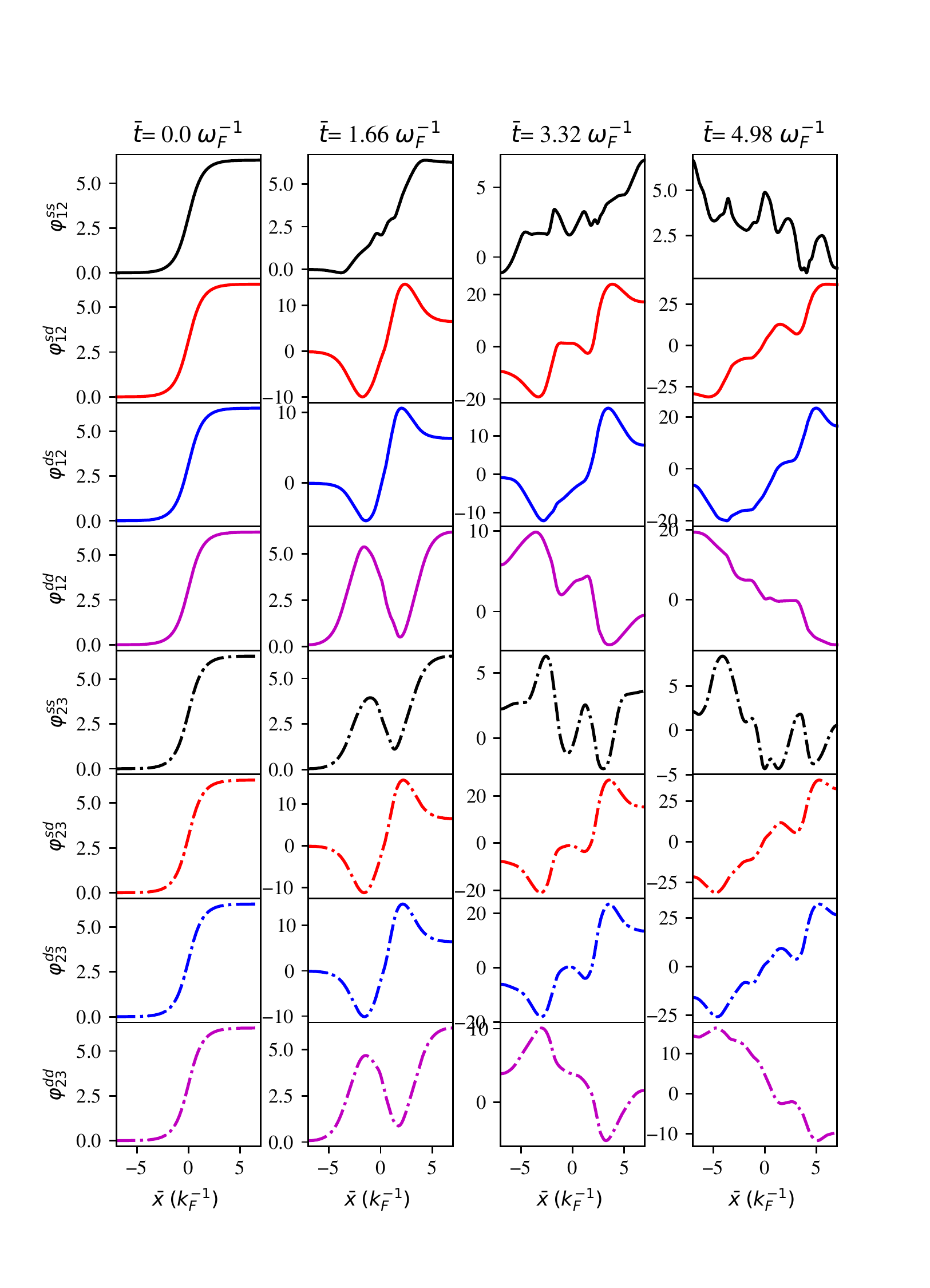}
	\caption{(Colour online) Spatial-temporal variation of phase differences in  various channels in the coupled LJJ with junction thickness of 6 {\AA} and layer thickness of 9 {\AA} under the application of bias voltage of 0.5 V.}\label{phase:03}
\end{figure}
\begin{figure}[!t]
\centering
	\includegraphics[clip,width=0.97\textwidth,angle=0]{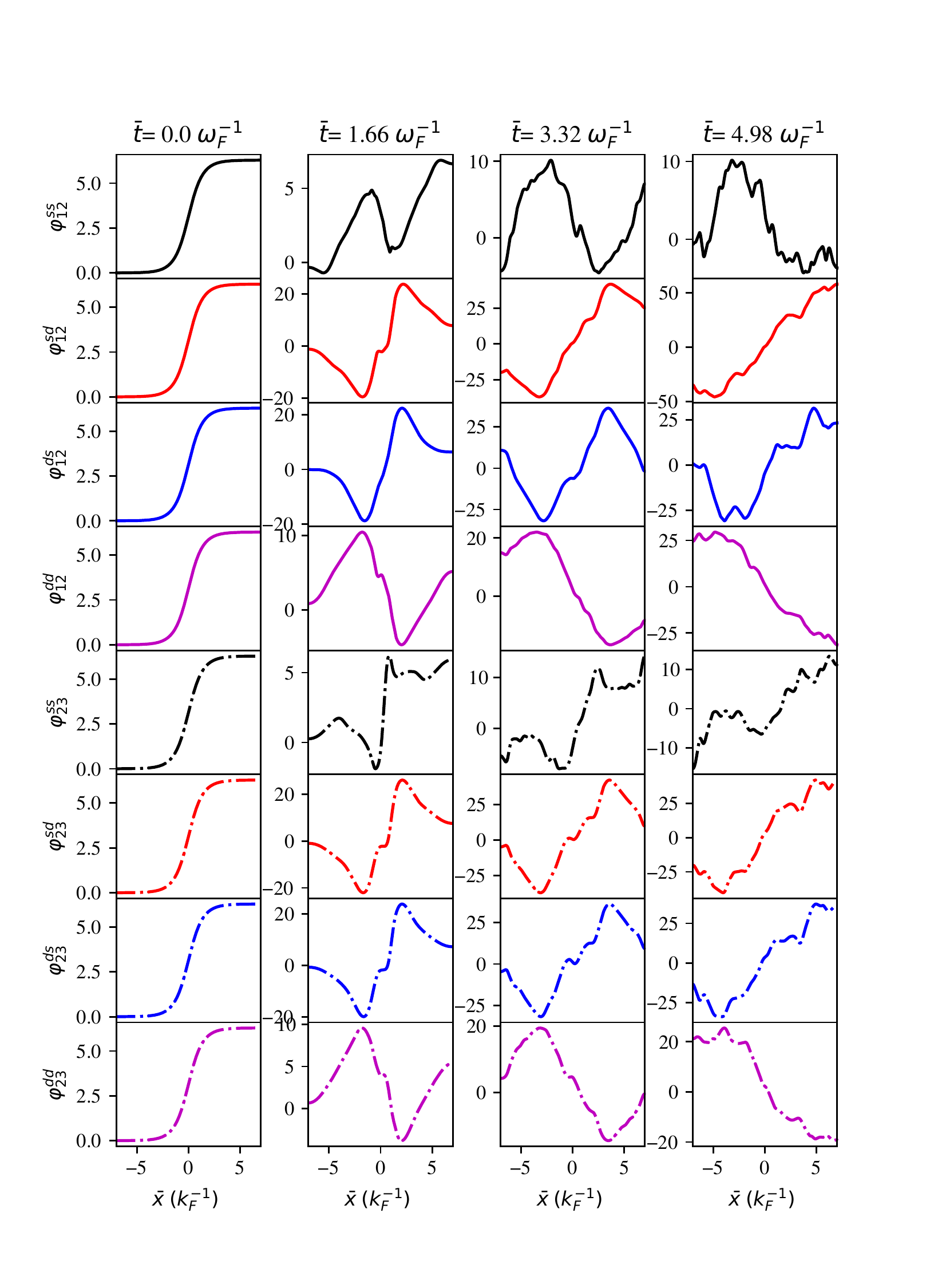}
	\caption{(Colour online) Spatial-temporal variation of phase differences in  various channels in the coupled LJJ with junction thickness of 6 {\AA} and layer thickness of 9 {\AA} under the application of bias voltage of 1 V.}\label{phase:04}
\end{figure}

There are different boundary conditions that can be imposed on the system in order to control the state of kink or anti-kink. When $\varphi(\bar{x},\bar{t})=0$ for $\bar{x}=\pm L_x$, then this condition will mirror the kink (anti-kink) and is known to be homogeneous Dirichlet boundary condition. The effect of moving $\varphi$  will be demonstrated at the boundary $\bar{x}=-L_x$ to feed the domain with the incoming kink(anti-kink). The boundary condition then reads 
\[ \varphi(-L_x,\bar{t})=4\tan^{-1}\left[\exp\left(\sigma\dfrac{-L_x-u\bar{t}-x_0}{\sqrt{1-u^2}}\right)\right]. \]
If the kink/anti-kink is to let reflecting from the boundary, then Neumann boundary condition, $\pard{\varphi}{\bar{x}}=0$ for $\bar{x}=\pm L_x$ can be used~\cite{langtangen2014}. 

In the present context, Neumann boundary condition is imposed which is approximated by central finite difference and yields 
\begin{gather}
\varphi_1^n=\varphi_{-1}^n,\quad \text{ at } \quad\bar{x}=-L_x, \quad\text{ and }\quad \varphi_{N_x+1}^n=\varphi_{N_x-1}^n,\quad \text{ at }\quad \bar{x}=+L_x.
\label{eq4.19}
\end{gather}

Putting $n=0$ in equation \eqref{eq4.09}  and using equation~\eqref{eq4.13},  we get
\begin{gather}
\varphi_i^{1}=\delta t\left.\pard{\varphi}{\bar{t}}\right|_{\bar{t}=0}+\varphi_i^0+\dfrac{\delta t^2}{2\delta x^2}\left(\varphi_{i+1}^0-2\varphi_i^0+\varphi_{i-1}^0\right)-\dfrac{1}{2}\delta t^2\mathcal{M}_{0d}^{-1}\mathcal{M}_{Fd}(\bar{j}\sin\varphi_i^n).
\label{eq4.25}
\end{gather}
The Courant-Friedrichs-Lewy stability criteria, $ \dfrac{\delta t^2}{\delta x^2}<1 $ should be maintained in order to get the stability of the kink/anti-kink solution.  
This criteria suggest us to take a very small time step as compared to the position step as far as possible. It is mandatory to pay the computational cost for the time steps to obtain the approximately calculated values, and reach a close agreement with those of theoretical values \cite{wang2014}. 

The most straightforward way to proceed the computational task is to introduce one array to hold $\varphi$  at all $\bar{x}_i$ at the time $\bar{t}_n$, a second array to hold all the $\varphi$  at the time $\bar{t}_{n-1}$ and a third array to hold a newly computed result at $\bar{t}_{n+1}$. Then, it is looped through the code incrementing the time and shuffling the arrays appropriately \cite{devries2011}.

The current can be calculated using \cite{koyama1996,koyama2008,koyama2010} 
\begin{equation}\label{curr}
I=\dfrac{\hbar\varepsilon_0c^2}{2e\lambda_F}\dfrac{\partial\varphi}{\partial \bar{x}}.
\end{equation} 
The current is averaged out over space and time as well as channel at different tunnel voltage which includes the element of equation \eqref{eq11.03} in the tunneling matrix. For the particular junction geometry, the tunneling matrix element is proportional to the bias voltage or tunnel voltage \cite{chen1990}. When the bias voltage is changed, then the tunnel matrix element also changes resulting in the change of the tunneling coupling constant of equation \eqref{eq11.03}. This variation in the tunneling coupling constant significantly contributes to the soliton motion represented by the phase differences in various channels.

\begin{figure}[!t]
	\begin{subfigure}[t]{0.49\columnwidth}
		\centering
		\includegraphics[width=\textwidth]{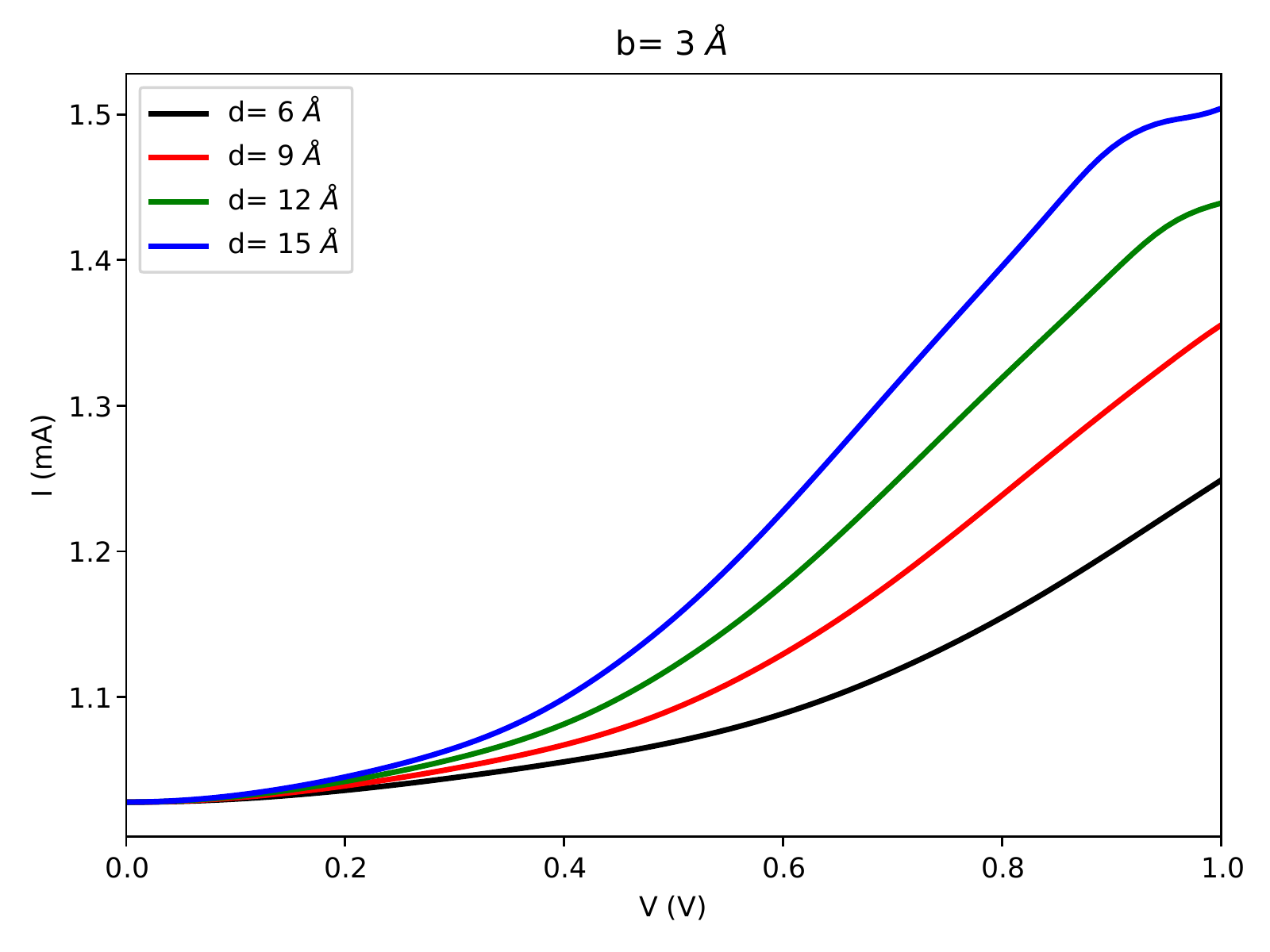}
		\caption{} \label{fig2a}
	\end{subfigure}
	\begin{subfigure}[t]{0.49\columnwidth}
		\centering
		\includegraphics[width=\textwidth]{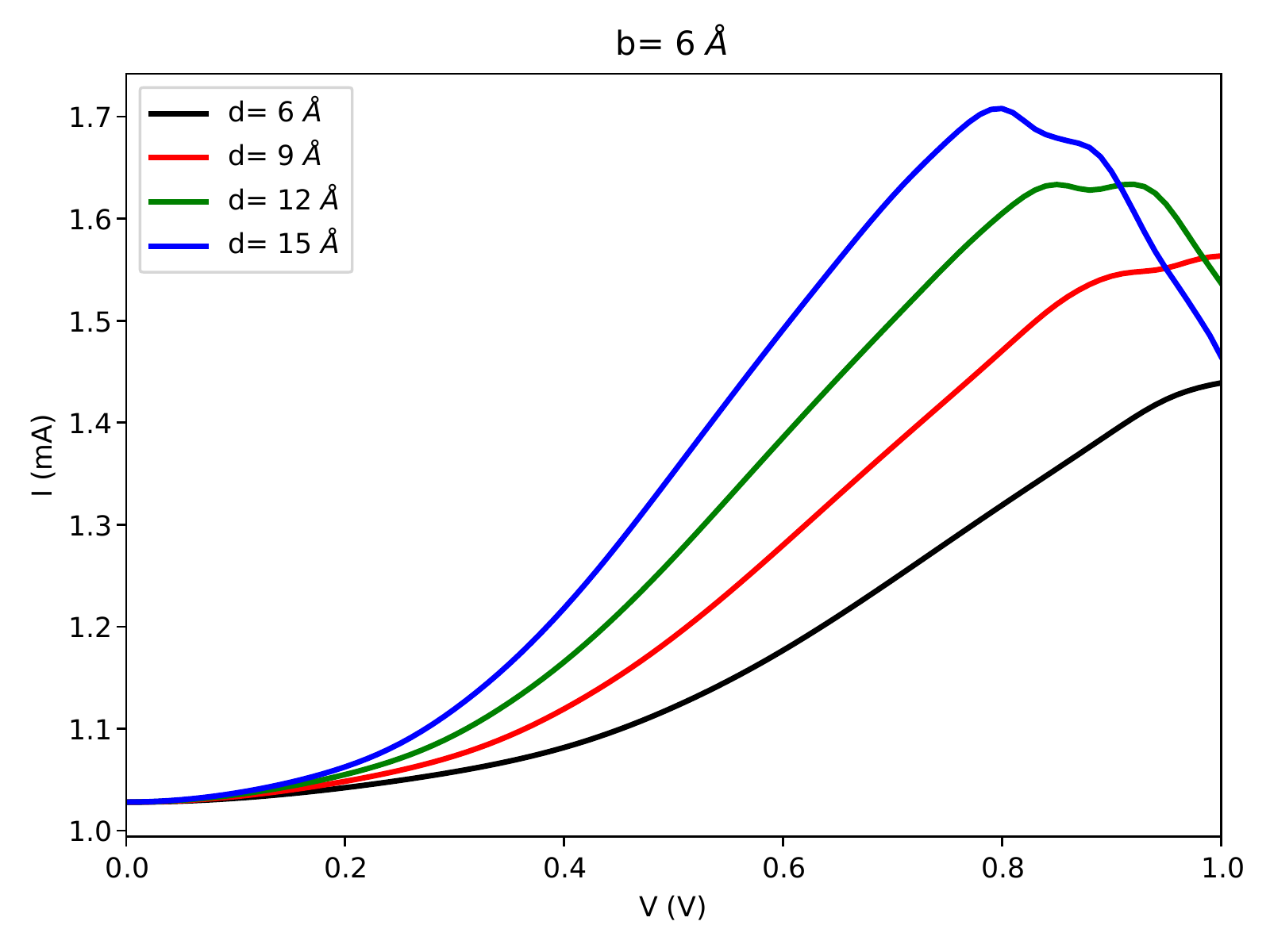}
		\caption{} \label{fig2b}
	\end{subfigure}\\
	\begin{subfigure}[t]{0.49\columnwidth}
		\centering
		\includegraphics[width=\textwidth]{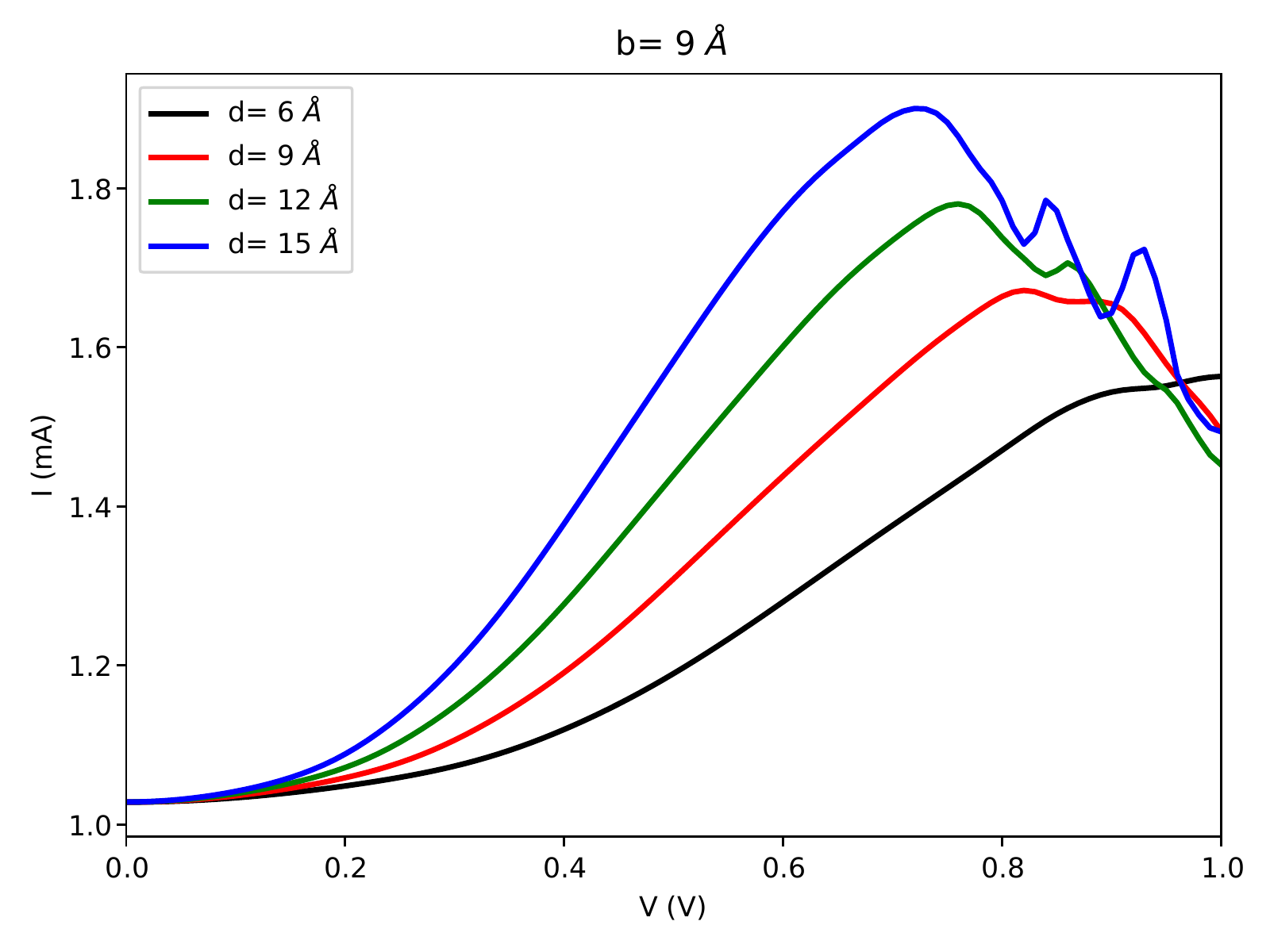}
		\caption{} \label{fig2c}
	\end{subfigure}
	\begin{subfigure}[t]{0.49\columnwidth}
		\centering
		\includegraphics[width=\textwidth]{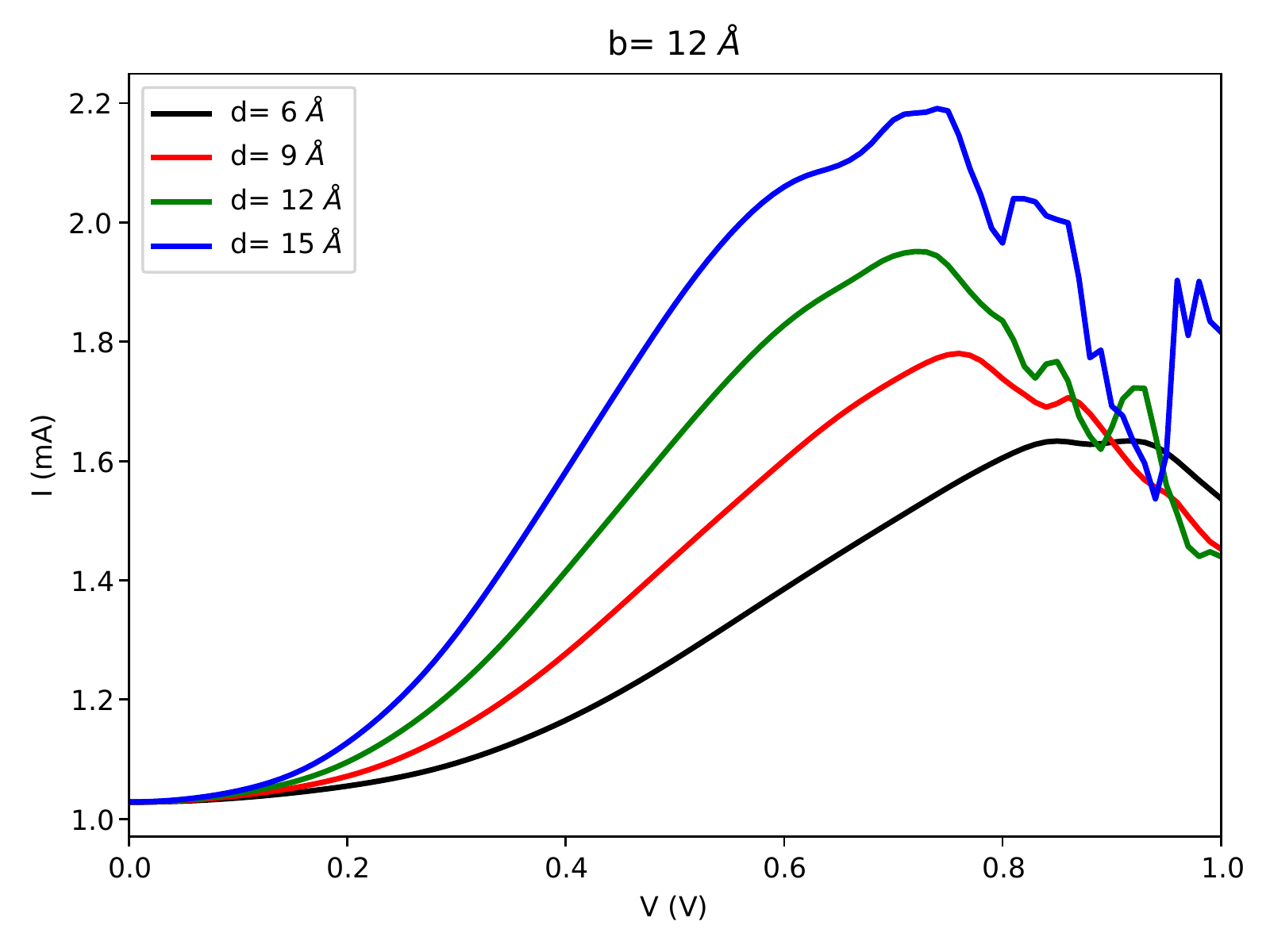}
		\caption{} \label{fig2d}
	\end{subfigure}
	\caption{(Colour online) Current-Voltage characteristics for junction thickness  (\subref{fig2a})  3~\AA,  (\subref{fig2b}) 6~\AA , (\subref{fig2c})  9~\AA , and (\subref{fig2d}) {12~\AA} for different layer thicknesses.}\label{fig2}
\end{figure}

The simulations were performed for a typical junction system of MgB$_2$ with superconducting layer thicknesses of {6\AA}, {9\AA}, {12\AA} and {15\AA} for different junction thicknesses {3\AA}, {6\AA}, {9\AA} and {12\AA} of SiO$_2$. The junction and superconducting layer thicknesses are taken in the range of molecular dimension (i.e., Angstrom) as suggested by Giaever \cite{giaever1960}. The dielectric constant of SiO$_2$ is taken as 3.7 and Fermi velocity of MgB$_2$ is taken as 4.7$\times$10$^5$ m/s \cite{buzea2001}. The computations were done using OCTAVE 4.4 programming language and the figures are generated using PYTHON3 in the Linux operating system on supercomputer platform. 

To study the plasmon excitation, the dispersion relation defined by equation \eqref{disp} is plotted and presented in  figure~\ref{plasma}. The figures \ref{plasma}\subref{plasma:a} to \ref{plasma}\subref{plasma:d} show that the band spectrum significantly depends on the junction and layer thicknesses. They appear in higher frequency states as the junction as well as the layer thicknesses are increased. The excited states can also be reached by increasing the bias-voltage that affects the tunneling matrix element. Hence, the tunneling coupling constant of equation \eqref{eq11.03} is also  altered.

In order to study the soliton motion, the phase differences $\varphi$ for all 8-channels are plotted against the normalized space and time. For this purpose, the LJJ of length 14 units was taken. The length is measured in the unit of inverse of Fermi wave vector (i.e. $\textbf{k}_\text{F}^{-1}$). The simulation was done up to the normalized time of 5 unit. Here, the time is measured in the inverse of Fermi frequency (i.e. $\omega_\text{F}^{-1}$). From figures \ref{phase:01} to \ref{phase:04}, it is observed that the initial kink in each channel greatly deforms its shape as the time lapses. As the kink reaches the boundary, it reflects back due to the application of Neumann boundary condition. As it reflects, there is a chance of producing the anti-kink or another kink. As a result, kink-kink or kink-anti-kink superposition may take place forming a complicated phase texture as time lapses. Some channels also show the collective behavior. The figures also show that the phase texture is highly sensitive to the junction and layer thicknesses as well as to the tunnel voltages. They become more complicated as the junction parameters are increased.

The I-V characteristics were studied by applying the voltage across the junction system. Since the tunneling matrix element is directly proportional to the bias voltage for the given junction and layer thicknesses \cite{chen1990}, the tunneling coupling constant and hence the phase differences for all channels were computed for different applied voltages. Using the equation \eqref{curr}, the current for each channel can be computed at the given applied voltage as the function of space and time. The current is then averaged out over space and time as well as the channels for the given applied voltage and junction geometry. In this way, the current is calculated for each applied voltage ranging from 0 V to 1 V with the step of 0.01 V. The current is plotted against voltage as shown in figures \ref{fig2}\subref{fig2a} to \ref{fig2}\subref{fig2d}. The I-V characteristics for the junction thickness of $b=3$ {\AA} are presented in figure \ref{fig2}\subref{fig2a}. The figure contains four curves for layer thicknesses of 6, 9, 12, and 15 {\AA}. These curves seem to of be of non-ohmic nature with the existence of differential resistance. This resistive nature indicates that the applied voltage was consumed in order to proceed the tunneling of Cooper pairs. As the Cooper pairs reach the junction, they break into the normal charge carrier and they collide with the lattice in the junction. The junction system behaves as a conventional resistor. The graph shows that the positive differential resistance increases as the increment of the layer thickness. As the layer thickness increases, the population of Cooper pair also increases. For this reason, the collision frequency of the carriers was increased and a greater resistance persisted for the same applied voltage.

When the junction thickness was changed to 6 {\AA}, an unusual type of I-V characteristics is obtained. For the given superconducting layer thickness, the I-V curve shows a positive differential resistance up to a certain applied voltage and then it shows a negative differential resistance. This peculiar behavior of the junction system indicates that there exists a complicated phenomenon. Due to the existence of a negative resistance at a certain voltage range, the device can be used as the  energy stroage  and oscillator. The origin of the noisy I-V characteristics may be interpreted in the following way. When the voltage is applied to the junction system, there are a lot of ways of dividing the value of this voltage into the voltages on those junctions. For this reason, there exist a lot of meta-stable states with different voltage distributions. Furthermore, it is quite possible that some meta-stable states are energetically very close to the stable state~\cite{koyama1996}. In such a case, the voltage distribution will be greatly changed and will cause a rapid oscillation of the dc current. For the junction thicknesses of 9 {\AA} and 12 {\AA}, the I-V curve showed even a complicated non-linear nature with the existence of a series of N-shaped differential resistances. The N-shaped differential resistance is the charanteristics of Gunn diode. Hence, the present junction system is also applicable to the low temperature electronic devices demanding the Gunn diode charactersitics. Another reason for the negative differential resistance is electromagnetic radiation due to the fluxon-antifluxon (kink-antikin) transition  between the plasmon excitation states during the soliton motion. As shown in the band spectrum (depicted in figure~\ref{plasma}), the plasma wave is at the excited state causing the plasmon radiation. Therefore, the junction system in a particular voltage range can be used as a radiation chamber.

\section{Conclusion}
We conclude that the collision of fluxon and anti-fluxon as well as the in-phase or the out-phase of collective motion is more active for higher tunnel voltage. The current voltage characteristics are almost linear up to a certain tunnel voltage and then become non-linear. The non-linearity starts at a lower tunnel voltage for higher junction thicknesses as well as layer thicknesses. The linear region indicates that the junction system demonstrates a resistive nature while the non-linear condition confirms the existence of other complicated phenomena. Some nonlinear regions confirm the existence of a negative differential resistance so that the the junction system can be used in the electric devices that demand a negative resistance. One of the phenomena is the emission of electromagnetic radiation (e.g., microwave, THz etc.) due to the formation of meta-stable states as predicted by Koyama \cite{koyama2010} and fluxon-antifluxon transition between  them. The device can be used as a switching device, a memory device that operates in  non-linear region. This might be the main region of THz radiation.

\section{Acknowledgement}
We would like to acknowledge our gratitude to our colleagues of Central Department of Physics, Tribhuvan University, Nepal for their valuable suggestions. We would also like to thank the Supercomputer team of Kathmandu University, Dhulikhel, Nepal for providing the computational facilities since otherwise the work could not be completed.

\ukrainianpart

\title{Теоретичне дослідження  I-V характеристик у зв'язаних довгих джозефсонівських переходах на основі надпровідника дибориду магнію}
\author{С.П. Чімоурія\refaddr{label1,label2},
        Б.Р. Гіміре\refaddr{label2}, Дж.Х. Кім\refaddr{label3}}
\addresses{
\addr{label1} Фізичний факультет, університет Катманду,Непал
\addr{label2} Фізичний факультет, Трибхуанський університет, Катманду, Непал
\addr{label3} Коледж природничих наук та інженерії, університет Х'юстона Кліер Лейк, TX, США
}

\makeukrtitle 
\begin{abstract}
У статті досліджено вольт-амперні (I-V)  характеристики у зв'язаному довгому джозефсонівському переході на основі дибориду магнію шляхом встановлення системи рівнянь різниці фаз різних інтер- та інтра-зонних каналів, починаючи з мікроскопічного гамільтоніана системи переходу та спрощення її за допомогою таких феноменологічних процедур, як дія, функція розподілу, перетворення Габарда-Стратоновича (бозонізація), інтеграл Грасмана, метод перевалу,  голдстоунівська мода, фазозалежний  ефективний лагранжіан і, нарешті, рівняння руху Ейлера-Лагранжа. Система рівнянь розв'язується з використанням скінченно-різницевого наближення, для якого за початкову умову приймається розв'язок незбуреного  синус-гордонівського рівняння. Гранична умова Неймана підтримується на обох кінцях так, що флаксон здатний відбиватися від кінця системи. Фазозалежний струм розраховується для різної тунельної напруги і усереднюється за простором і часом. Вольт-амперні характеристики майже лінійні при низькій напрузі та нелінійні при більш високій напрузі, що вказує на те, що в цій ситуації можуть виникнути більш складні фізичні явища. У деяких областях характеристик існує негативний опір, що означає, що система переходів може бути використана в певних електронних пристроях, таких як генератори, перемикачі, пристрої пам'яті тощо. Нелінійність також чутлива до шару, а також до переходу товщини. Нелінійність виникає при меншій напрузі, а також при більшій товщині переходів і шарів.

\keywords двозонний надпровідник, зв'язаний довгий джозефсонівський перехід, перетворення Габарда-Стратоновича, збурене рівняння синус-Гордона
\end{abstract}

\lastpage

\end{document}